\documentclass[10pt,a4paper]{book}
\usepackage{udineHyderabad-Klu2}

\begin{document}

\talktitle{Contribution of pulsars to the gamma-ray background and their
observation with the space telescopes GLAST and AGILE}{Contribution of pulsars to the gamma-ray background and their
observation with the space telescopes GLAST and AGILE}

\talkauthors{Erica Bisesi\structure{a}}

\authorstucture[a]{Department of Physics, 
                   Udine University, 
                   via delle Scien\-ze~208, 33100~Udine, Italy}

\shorttitle{Contribution of pulsars to the gamma-ray background} 

\firstauthor{E.\ Bisesi}

		\index{Bisesi@\textsc{Bisesi}, E.}

\begin{abstract}Luminosities and fluxes of the expected population of galactic gamma-ray pulsars become foreseeable if physical distributions at birth and evolutive history are assigned. In this work we estimate the contribution of pulsar fluxes to the gamma-ray background, which has been measured by the EGRET experiment on board of the CGRO. For pulsar luminosities we select some of the most important gamma-ray emission models, taking into account both polar cap and outer gap scenarios. We find that this contribution strongly depends upon controversial neutron star birth properties. A comparison between our simulation results and EGRET data is presented for each model, finding an average contribution of about 10\%. In addition, we perform the calculation of the number of new gamma-ray pulsars detectable by GLAST and AGILE, showing a remarkable difference between the two classes of models. Finally, we suggest some improvements in the numerical code, including more sophisticated galactic m odels and different populations of pulsars like binaries, milliseconds, anomalous pulsars and magnetars.%
\end{abstract}

\section{Introduction}
The estimation of the contribution of pulsars to the gamma-ray background is
based on the comparison between theoretical predictions on the flux emitted by
these sources and experimental observations. Each model for the
emission of radiation by a stellar population is specified by the galactic
distribution of sources and their individual luminosities.
\vspace*{-.5mm}

\section{Galactic model}
\subsection{Spatial distributions}
\begin{figure}[t]
\centering
\includegraphics[width=9cm]{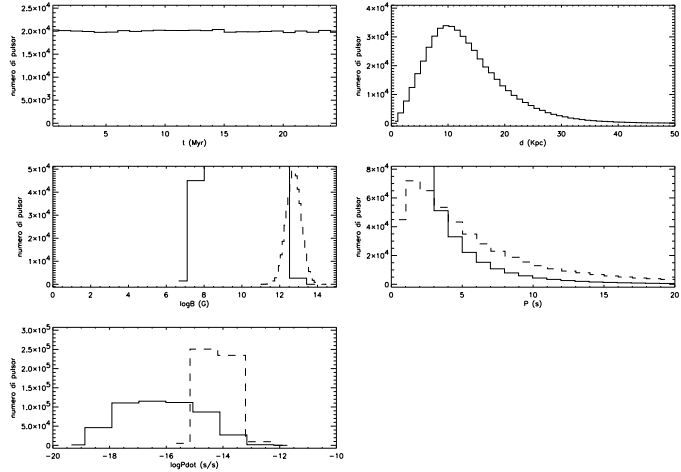}
\caption{Distributions of pulsars ages $(t)$, distances $(d)$, magnetic fields
$(B_0)$, periods $(P_0)$, period derivatives $(\dot{P})$. Dot lines represent
a situation with constant magnetic field, solid lines are referred to the
field decay case (E.\ Bisesi, 2002).}
\label{Fig.1-bis}
\end{figure}

We here present a numerical simulation of the galactic population of radio
pulsars giving all initial spatial and physical parameters in the framework
of the galactic models of Paczy\'nski (1990) and of Gonthier \& al.\ (2002) 
\protect\cite{4-bis}; we then make them evolve accordingly with the velocity model of 
Sturner \& Dermer (1994) \protect\cite{10-bis}.
We show histograms of distributions of various pulsar properties in Fig.~(\ref{Fig.1-bis}), assuming both constant magnetic field and field decay.

\subsection{Pulsar luminosities}
In the commonly accepted hypotesis that every pulsar irradiates on the whole
electromagnetic spectrum, we assign a gamma luminosity to each source. We
consider a suitable choise of emission models. \vspace*{\baselineskip}

\textbf{POLAR CAP MODELS:}

\begin{itemize}
\item \textbf{Harding (1981) \protect\cite{5-bis}:} $L_{\gamma} \mbox{($>$\mbox{100 MeV})} = 1.2
  \times 10^{35} B_{12}^{0.95}\, P^{-1.7}\, \mbox{ ph}\, \mbox{ s$^{-1}$}$;
\item \textbf{Zhang \& Harding (2000) \protect\cite{12-bis}:} 
\item [] \hspace{1cm} $L_{\gamma}(I) = 5.87 \times 10^{35} {B_{12}}^{6/7}\,
  P^{-1/7} \mbox{ ph} \mbox{ s$^{-1}$} \hspace{0.3cm} \mbox{ for} \hspace{0.3cm} {B_{12}}^{1/7}\, P^{-11/28}
  > 6.0,$
\item [] \hspace{1cm} $L_{\gamma}(II) = 1.0 \times 10^{35} {B_{12}}\,
  P^{-9/4} \mbox{ ph} \mbox{ s$^{-1}$} \hspace{0.3cm} \mbox{ for}
  \hspace{0.3cm} {B_{12}}^{1/7}\, P^{-11/28} < 6.0;$
  
\begin{figure}[t]
\centering
\includegraphics[width=6cm]{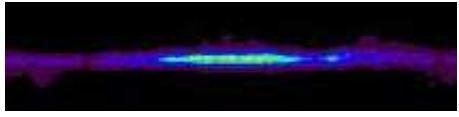}
\caption{Map\label{Fig.2-bis} of the gamma-ray sky as seen by EGRET (E.\ Bisesi, 2002).}
\end{figure}

\begin{figure}[ht]
\centering
\includegraphics[width=9cm]{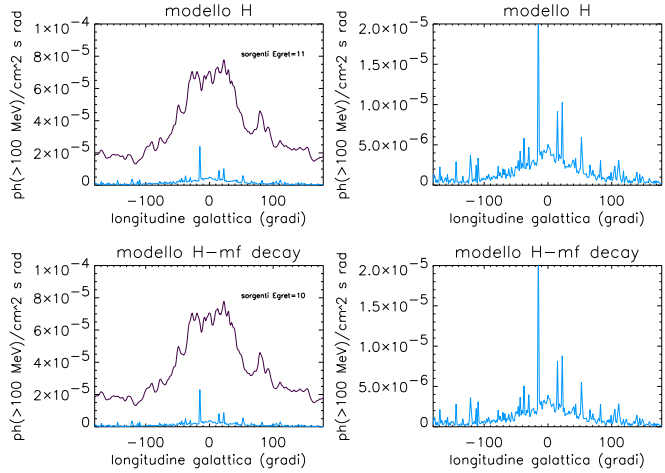}
\caption{Comparison\label{Fig.3-bis} between the EGRET gamma-ray background (in blue) and
predictions for the polar cap model of Harding (1981) (in violet). We also give
the simulated number of pulsars above the EGRET detection threshold for this
model. Plots on the right side are enlargements of the simulated profiles (E.\ Bisesi, 2002).}
\end{figure}

\item \textbf{Sturner \& Dermer (1994) \protect\cite{10-bis}:} $L_{\gamma} = 6.25 \times 10^{35} {B_{12}}^{3/2}\, P^{-2} \mbox{ ph} \mbox{ s$^{-1}$}.$
\end{itemize}
\vspace{0.5cm}

\textbf{OUTER GAP MODELS} with the \emph{death line} $5\, \log{B} - 12\, \log{P} \leq 72$:

\begin{itemize}
\item \textbf{Romani \& Yadigaroglu (1995) \protect\cite{9-bis}:} $L_{\gamma}\! =\! 1.56 \times
  10^{36} {B_{12}}^{0.48}\, P^{-2.48} \mbox{ ph}\! \mbox{ s$^{-1}$};$
\item \textbf{Cheng \& Zhang (1996) \protect\cite{2-bis}:} $L_{\gamma} = 3.93 \times 10^{37} {B_{12}}^{0.3}\, P^{-0.3} \mbox{ ph} \mbox{ s$^{-1}$}.$
\end{itemize}

We calculate the integrated flux on the galactic latitude  $-10^{\circ}
< b < 10^{\circ}$ as a function of the longitude $l$; results are
plotted together with the experimental points as they have been measured by
the EGRET experiment.

Fig.~(\ref{Fig.2-bis}) and (\ref{Fig.3-bis}) respectively show the map of the gamma-ray sky as seen by EGRET and our results for the polar cap model of Harding (1981).

%

The contribution of pulsars to the gamma-ray background is less than 10\% for
this model; the foreseen number of sources above the EGRET detection threshold of
$10^{-7}$ ph cm$^{-2}$ s$^{-1}$  in not far from the observed value.

\section{Comparison among models}
In the previous discussion we have forced the physical parameters involved in
the calculation of gamma ray luminositied and fluxes to assume 
some definite values. In order to give a 
meaningful comparison among different emission models an extension to a more 
general parameter space is required.

So we define a parameter space whose extremes are corresponding to the
limit conditions for pulsating gamma neutron stars; we then follow the
evolution of a representative point inside this space. The most relevant
physical parameters are the rotation period, the magnetic field and the
velocity fron the Galactic Centre. 

\subsection{Pulsar populations}
\begin{figure}[!t]
\centering
\includegraphics[width=6cm]{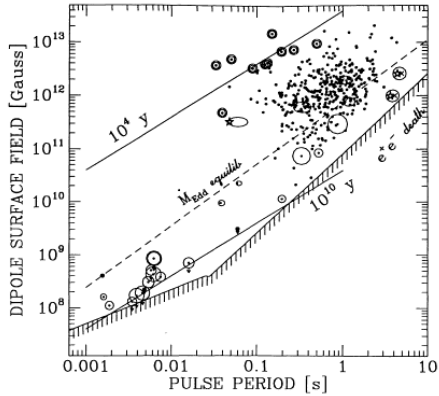}
\caption{Diagram period$-$magnetic field for different pulsar
families. Points indicate young isolated pulsars, circle$-$dots pulsars in
binary systems. The population of millisecond pulsars is visible at the bottom
left (from Phinney \& Kulkarni, 1994 \protect\cite{8-bis}).}
\label{Fig.4-bis}
\end{figure}

Different families of pulsars follow distinct evolutive histories, and this
has a direct impact on their physical properties. Furthemore, physical 
mechanisms ruling over the gamma-ray emission are peculiar of each class. 
Fig.~(\ref{Fig.4-bis}) shows a diagram period$-$magnetic field for young isolated, 
binaries and millisecond pulsars. Magnetars are characterized by very high 
magnetic fields, typically $10^{14}-10^{16}$ G.

We bind the parametric space in this way:
\begin{itemize}
\item $P_{0min} = 0.01$ s (when the centrifugal force equals molecular bonds
of the star, yielding its survival);
\item $P_{0max} = 1$ s (fixed from observations);
\item $B_{0min} = 10^{11}$ G (the lowermost value for young isolated pulsars);
\item $B_{0max} = 4.413 \times 10^{13}$ G (condition for the magnetar regime);
\item $v_{0min} = 120$ km/s (Gonthier \& al., 2002 \protect\cite{4-bis});
\item $v_{0max} = 350$ km/s (Lyne \& Lorimer, 1994).
\end{itemize}

\begin{figure}[!t]
\centering
\includegraphics[width=6cm]{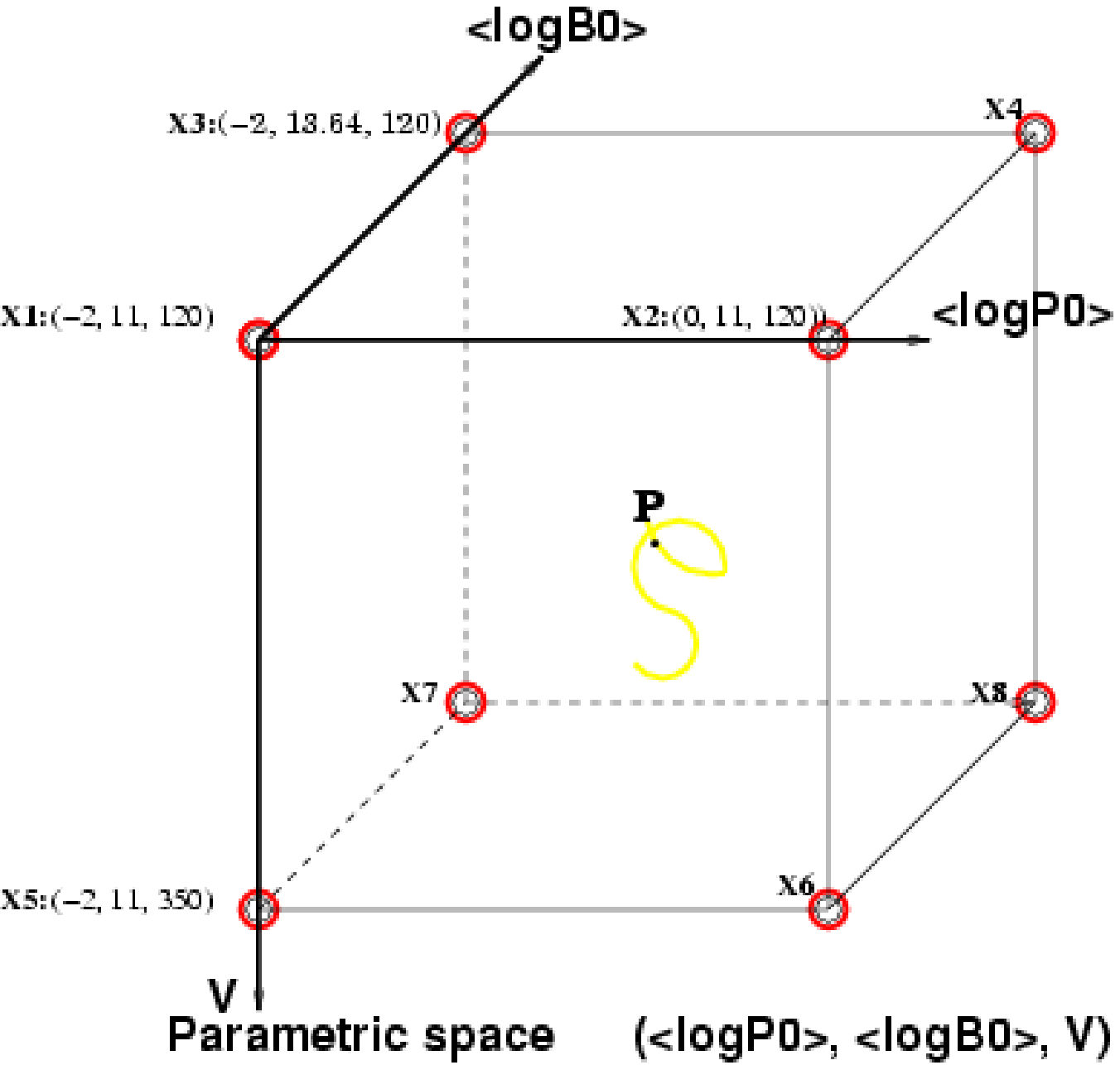}
\includegraphics[width=6cm]{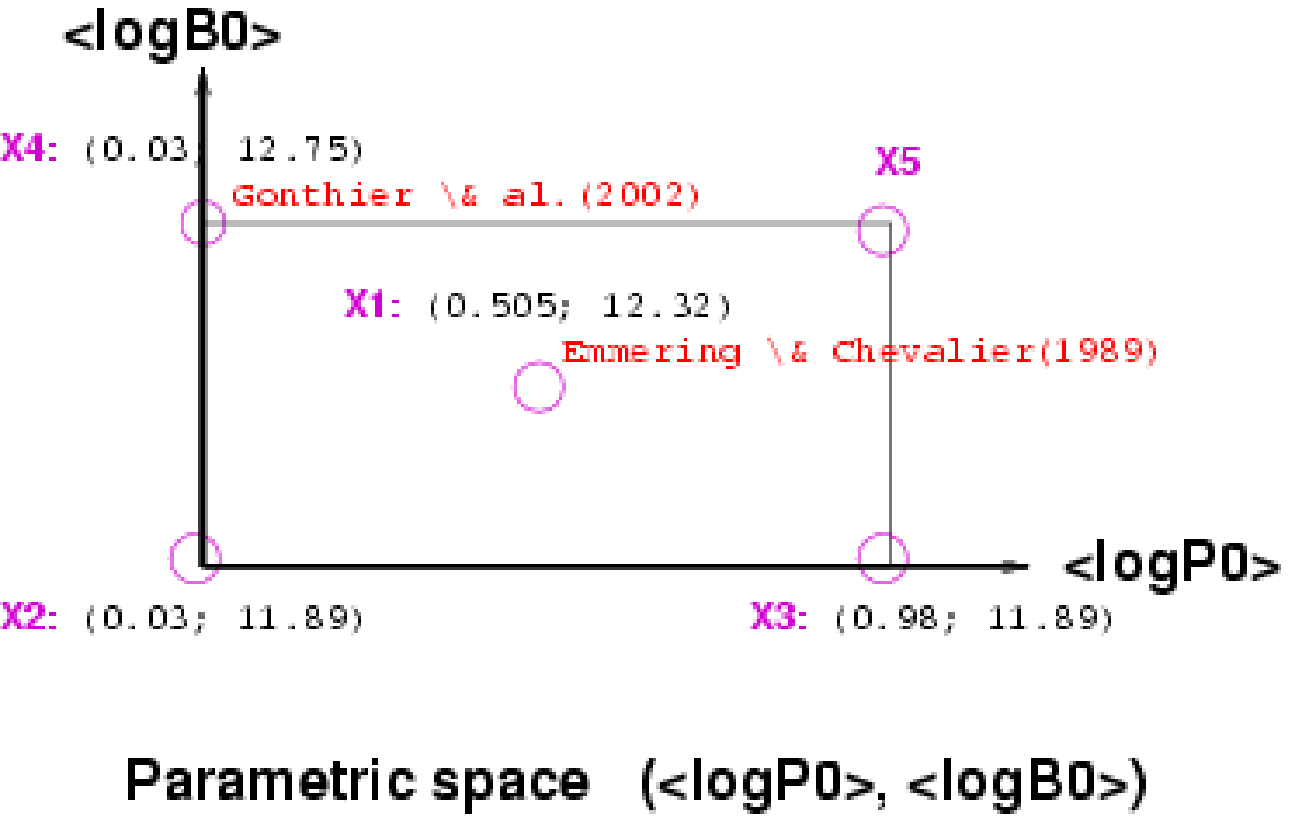}
\caption{Comparison among models (E.\ Bisesi, 2002).}
\label{Fig.5-bis}
\end{figure}

We show a tridimensional representation of the parametric space
$\langle\log{P_{0}}\rangle$, $\langle\log{B_{0}}\rangle$, $V$ in the left side
of Fig.~(\ref{Fig.5-bis}). The bidimensional portion on the right shows the
positions of two models of interest \protect\cite{4-bis,3-bis}.

\begin{figure}[!ht]
\centering
\includegraphics[width=7cm]{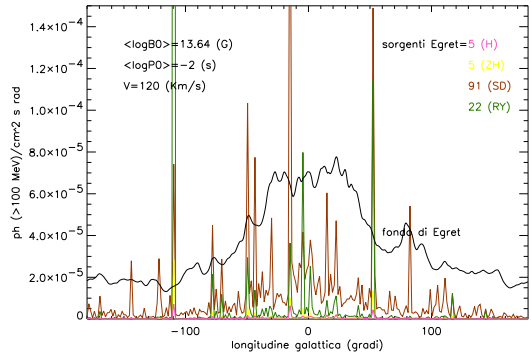}
\caption{Comparison between EGRET observations and simulation results for the
point $X_{3}$ of the parametric space (E.\ Bisesi, 2002).}
\label{Fig.6-bis}
\end{figure}

We compare predictions for the five models introduced above for a set of
points of the parametric space. As physical mechanisms for the gamma-ray 
emission differ from one pulsar family to another, we would give luminosity
function expressions for each of them. We purpose to improve our work in such
a way in future; in this context we restrict our analysis to young isolated 
pulsars. An example relevant to the point $X_{3}$ is shown in 
Fig.~(\ref{Fig.6-bis}).

\section{Discussion}
For each point considered, we evaluate the percentual excess between the
foreseen number of gamma-ray pulsars for each model and their actual number of
7. The total percentual excess for each model gives us the possibility of 
select the most reliable model, which we find being that of Harding (1981)
\protect\cite{5-bis}.

We finally estimate the number of gamma-ray pulsars detectable by the next
space telescopes GLAST and AGILE for the model selected.

GLAST and AGILE detection thresholds are $6 \times 10^{-9}$ ph cm$^{-2}$
s$^{-1}$ and $10^{-7}$ ph cm$^{-2}$ s$^{-1}$ respectively.

Our results for the five points of the right side of Fig.~(\ref{Fig.5-bis}) are 
shown in Table 1.

Predictions for GLAST are very optimistic, remarkably we expect to detect a
very large number of new gamma-ray pulsars, opening very promising frontiers
in understanding these mysterious and fascinating objects.

\begin{table}[!hb]
\centering
\caption{Total percentual excess for each model on the
whole parametric space for young pulsars.}
\scriptsize
\begin{tabular}{c c } 
\multicolumn{2}{c}{}\\\hline\hline
\footnotesize{Model} & \footnotesize{$|$Total percentual excess$| =
\sum_{i=1}^{8}|$Percentual excess($X_{i})|$}\\
\hline
$H$ & $27$\\
$ZH$ & $220$\\
$SD$ & $174$\\
$RY$ & $30506$\\
$CZ$ & $2277$\\
\hline
\end{tabular}
\end{table}

\end{document}